\documentclass[aps,prl,twocolumn,superscriptaddress,groupedaddress,nofootinbib]{revtex4-1}  % for review and submission
\usepackage{graphicx}  % needed for figures
\usepackage{dcolumn}   % needed for some tables
\usepackage{bm}        % for math
\usepackage{amssymb}   % for math
\usepackage{hyperref}

\begin{document}

\title{Transit Fare Arbitrage: Case Study of San Francisco Bay Area Rapid Transit (BART) System}
\author{\href{mailto:asifhq@gmail.com}{Asif Haque}}
\affiliation{\href{http://twitter.com/haqueasif}{Data Scientist, Twitter Inc.}}

\begin{abstract}
Transit fare arbitrage is the scenario when two or more commuters agree to swap tickets during travel in such a way that total cost is lower than otherwise. Such arbitrage allows pricing inefficiencies to be explored and exploited, leading to improved pricing models. In this paper we discuss the basics of fare arbitrage through an intuitive pricing framework involving population density. We then analyze the San Francisco Bay Area Rapid Transit (BART) system to understand underlying inefficiencies. We also provide source code and comprehensive list of pairs of trips with significant arbitrage gain at \href{https://github.com/asifhaque/transit-arbitrage}{\texttt{github.com/asifhaque/transit-arbitrage}}. Finally, we point towards a uniform payment interface for different kinds of transit systems.
\end{abstract}

\maketitle

\section{Motivation}
\label{sec:motivation}

With ever increasing global population and even faster densification of  various large metropolitan areas of the world, mass transit systems are becoming more important. Improved transit systems are also important for keeping personal cars off of congested road networks, thus addressing climate change concerns and making future switch to cleaner energy source easier. Urban planners are hard at work designing faster, cheaper, safer and more comfortable transit systems.

Most often local and federal governments are in charge of building transit systems. If we view transit systems as a public utility then this government control is reasonable. But it also means that there is almost no competition in most cities. It is widely believed that for most businesses competition may improve efficiency and quality of service. One area where fierce competition has helped reduce inefficiencies is electronic capital markets. Global stock exchanges and market places are more at sync and bid-ask spreads are low. One core principle that guides such optimization is the elimination of arbitrage or riskless profit making.

The goal of this paper is to investigate whether there are pricing inefficiencies in mass transit systems that can lead to arbitrage and ask how we can improve transit fares for commuters. As a case study we looked at San Francisco Bay Area Rapid Transit (BART) system, which is the fifth largest transit system by ridership in the US. We also took a cursory look at Washington DC Metrorail, which is the second largest system in the US.

A motivating example is the case when one individual is traveling from Millbrae station, a suburb south of San Francisco, to Embarcadero station which is downtown San Francisco, and at the same time a second individual is traveling from Glen Park station, a residential area in San Francisco, to Berkeley station where University of California Berkeley is located. The two tickets cost (according to \href{http://www.bart.gov/sites/default/files/docs/BART\%202014\%20Fare\%20Chart_Tab.pdf}{2014 fare chart}) \$4.50 and \$4.20 respectively. But if during the segment between Glen Park station and Embarcadero station the two travelers agree to exchange their tickets the cost becomes \$5.10 and \$1.85. So from a total cost of \$8.70 a simple ticket swap saves \$1.75 or 20\%.

Another similar example comes from Washington DC Metrorail. If one individual is traveling from Vienna/Fairfax station to Metro Center station on the Orange line and another individual is traveling from Rosslyn station to New Carrollton station on the same train then during peak hours they pay \$5.3 and \$4.9. But if the travelers decide to swap tickets they pay \$5.75 and \$2.1 instead. So the saving is \$2.35 or 23\% of total trip cost.

In the following sections we discuss general conditions for arbitrage and how those are reflected in real systems like BART.

\section{Generalization}
\label{sec:generalization}

In this paper we use the phrase \emph{fare arbitrage} as the scenario where two individuals traveling separately have overlapping routes and may agree to exchange tickets to decrease the total price of their trips. Higher order arbitrage where more than two travelers decide to exchange tickets is not considered here.

\begin{figure}
\includegraphics[width=\linewidth]{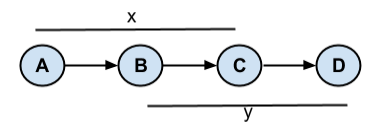}
\caption{Fare arbitrage scenario involving two commuters with overlapping paths. One commuter is going from station A to C while the other one is going from B to D.}
\label{fig:transit-diagram}
\end{figure}

Fig~\ref{fig:transit-diagram} shows a transit route from station A to D via stations B and C. Suppose one traveler is going from station A to C while another traveler is going from station B to D. Between stations B and C both travelers are on the same train. If they choose to swap tickets the transit system will see one traveler going from station A to D while another from B to C. If the total price of latter two tickets is less than the actual trips then there is fare arbitrage.

The reverse case could also be true -- if two travelers going from A to D and B to C respectively have to pay higher price than A to C and B to D then arbitrage becomes possible.

One simple pricing method that is quite common is to have a fixed fare for all trips. Large transit systems such as New York City and Chicago have such flat rates. Although this method eliminates fare arbitrage, it may only be feasible for a city with a large population, high density and high fraction of citizens using the service for daily commute.

\begin{figure}
\includegraphics[width=\linewidth]{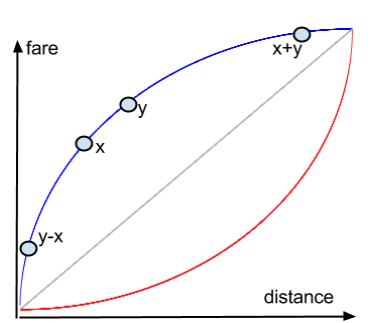}
\caption{Fare as a function of distance traveled. Center grey line is the price proportional to distance. The blue line above the grey line penalizes short distances while the red line below subsidizes them. Fig~\ref{fig:fare-scenario} attempts to explain these curves using population density.}
\label{fig:dist-vs-fare}
\end{figure}

Another common strategy is to use a price proportional to distance traveled. This fair pricing model is showed as the grey line in fig~\ref{fig:dist-vs-fare}. San Francisco Bay Area CalTrain system breaks up the route into zones and uses prices that are proportional to the number of zones traveled. Like the flat rate method, this pricing strategy also eliminates fare arbitrage. But it may not be optimal for revenue maximization if the transit system is complex and population density is non-uniform. It may also not be ideal if policy makers wish to change population density by encouraging people to move closer to city center or farther out.

Going back to the example in fig~\ref{fig:transit-diagram} let us assume that the individual going from station A to C is traveling distance $x$ while the individual going from station B to D is traveling distance $y$. Suppose $y > x$ and $d = y-x$. Let us assume $f(.)$ is the fare as a function of distance. We can express the arbitrage condition as follows.
$$ f(x) + f(y) \ne f(x+y) + f(d) $$

If $f(x) +f(y) > f(x+y) + f(d)$ and our two travelers swap their tickets during their trip, they get a discount worth $f(x) + f(y) - f(x+y) - f(d)$. This scenario is possible if the fare function has a concave shape, as shown by the blue curve in fig~\ref{fig:dist-vs-fare}.

In fig~\ref{fig:dist-vs-fare} the blue curve is above the fair grey line. This implies short distances are penalized to maxmize transit revenue. If the blue line had the same concave shape but were below the grey line then short distances are fair but long distances are subsidized.

If $f(x) + f(y) < f(x+y) + f(d)$ then a pair of travelers where one is traveling from A to D and the other from B to C can get a discount of $f(x+y) + f(d) - f(x) - f(y)$ if they choose to swap tickets mid trip. Convex price functions such as the red curve in fig~\ref{fig:dist-vs-fare} give rise to such arbitrage.

Since the red curve in fig~\ref{fig:dist-vs-fare} is below the grey line, short distances are subsidized but long distances are fair. A similar convex curve above the grey line would mean short distances are fair but long distances are penalized.

\begin{figure}
\includegraphics[width=\linewidth]{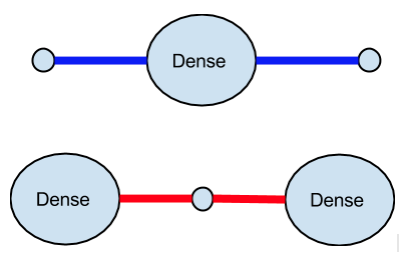}
\caption{Population density scenario for mass transit systems. Top diagram shows a (blue) transit line passing through a dense region while the bottom diagram shows a (red) transit line connecting two dense regions.}
\label{fig:fare-scenario}
\end{figure}

Hypothetical population density scenarios related to price functions in fig~\ref{fig:dist-vs-fare} are shown in fig~\ref{fig:fare-scenario}. In the top diagram a (blue) transit line starts and ends in sparse suburbs and passes through a dense region. In this case revenue can be increased by following the blue concave curve in fig~\ref{fig:dist-vs-fare} where there is an extra charge to enter the dense region.

In the bottom diagram of fig~\ref{fig:fare-scenario} the (red) transit line connects two dense regions, with sparse area in between. In this case revenue is maximized if fares follow a convex curve similar to the red curve in fig~\ref{fig:dist-vs-fare} but above the grey fair price line.

Both these scenarios are likely in real world and both give rise to fare arbitrage. Intuitively, the top diagram in fig~\ref{fig:fare-scenario} should match global structure of most metropolitan areas. San Francisco BART and Washington DC Metrorail does indeed resemble this. But we also see evidence of smaller instances of the bottom diagram of fig~\ref{fig:fare-scenario} in some part of the transit system.

\section{Case Study of BART}
\label{sec:casestudy}

\begin{figure}
\includegraphics[width=\linewidth]{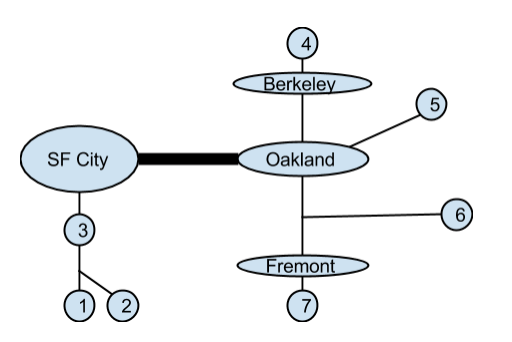}
\caption{Simplified map of Bay Area Rapid Transit (BART) system with dense regions shown in ovals. Numbered small circles are starting stations of various routes. Routes are listed in table~\ref{tab:bart-routes}. The thick line between San Francisco city and Oakland represents the bay tunnel.}
\label{fig:bart-map}
\end{figure}

San Francisco Bay Area Rapid Transit (BART) is the fifth largest transit system in the US with almost half a million riders on average on a weekday. Fig~\ref{fig:bart-map} shows a simplified diagram of BART with denser areas in ovals.

Five routes of BART according to fig~\ref{fig:bart-map} are shown in table~\ref{tab:bart-routes}. Four out of the five routes pass through dense San Francisco city, cross the bay tunnel (thick line in fig~\ref{fig:bart-map}), then through dense Oakland and end in suburban areas. The other route is the Richmond-Fremont line which passes through Oakland. All five routes resemble the top diagram of population scenario in fig~\ref{fig:fare-scenario}.

\begin{figure}
\includegraphics[width=\linewidth]{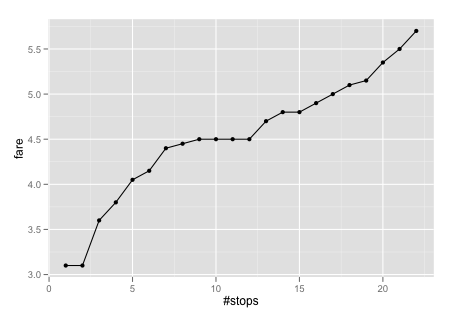}
\caption{Fare as a function of number of stops from Millbrae on Millbrae-Richmond line.}
\label{fig:millbrae-richmond}
\end{figure}

Fig~\ref{fig:millbrae-richmond} shows fares from Millbrae towards Richmond at various stations as a function of number of stops. We notice that the region between station 3 and station 7 corresponds to the steep price of entering San Francisco city. Once inside downtown San Francisco fares from Millbrae are flat until the bay tunnel is crossed. There is a small bump between station 12 and station 14 to enter Oakland. Beyond Oakland fare increases roughly linearly.

\begin{table}
\caption{BART routes in fig~\ref{fig:bart-map}.}
\begin{tabular}{|c||r|}
\hline
Route & Between stations \\ \hline \hline
Millbrae-Richmond & 1 and 4 \\ \hline
SFO Airport-Pittsburg & 2 and 5 \\ \hline
Daly City-Dublin & 3 and 5 \\ \hline
Daly City-Fremont & 3 and 6 \\ \hline
Richmond-Fremont & 4 and 7 \\ \hline
\end{tabular}
\label{tab:bart-routes}
\end{table}

The route from San Francico airport has a higher fare, presumably due to airport fees. But we see the same fare curve for the four routes through San Francisco city. The steep parts of fig~\ref{fig:millbrae-richmond} correspond to the blue concave curve in fig~\ref{fig:dist-vs-fare} which, as we have established earlier, allows arbitrage. The linear part beyond city centers corresponds to the fair grey line in fig~\ref{fig:dist-vs-fare}. In fig~\ref{fig:millbrae-richmond} there is a small convex section near the end of route between station 17 and station 22 resembling the red curve in fig~\ref{fig:dist-vs-fare}.

\begin{figure}
\includegraphics[width=\linewidth]{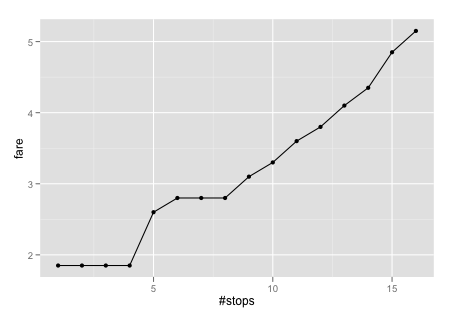}
\caption{Fare as a function of number of stops from Richmond on Richmond-Fremont line.}
\label{fig:richmond-fremont}
\end{figure}

The Richmond-Fremont route fares from Richmond are shown in fig~\ref{fig:richmond-fremont}. The bump corresponding to Oakland is smaller and fare is linear beyond Oakland. There is another tiny convex section near the end of the route.

Although arbitrage condition, as laid out earlier, is true for the concave bumps in fig~\ref{fig:millbrae-richmond} and fig~\ref{fig:richmond-fremont}, actual arbitrage becomes possible for one more reason -- price asymmetry. Fares from different stations even on the same route are different and unless these are carefully aligned it would be possible to pay less by swapping tickets. Table~\ref{tab:bart-arbitrage} shows a sample of pairs of routes where fare arbitrage is possible.

\begin{table}
\caption{Sample BART routes where fare arbitrage is possible. Gain as a percentage of total trip cost is shown in parentheses.}
\begin{tabular}{|c|c|l|}
\hline
Route 1 & Route 2 & Gain (\%) \\ \hline \hline
sf-airport $\to$ embarcadero & balboa-park $\to$ fremont & 1.80 (12) \\ \hline
millbrae $\to$ montgomery-st & balboa-park $\to$ walnut-creek & 1.80 (18) \\ \hline \hline
richmond $\to$ fremont & bay-fair $\to$ hayward & 1.05 (15) \\ \hline
orinda $\to$ pittsburg & walnut-creek $\to$ concord & 2.05 (36) \\ \hline
\end{tabular}
\label{tab:bart-arbitrage}
\end{table}

The first two rows in table~\ref{tab:bart-arbitrage} are due to the higher price commuters have to pay when they travel from San Francisco to Oakland compared to commuters who have started outside San Francisco and already paid a high price to enter the city. Someone going from San Francisco airport to Embarcadero can swap tickets with someone from Balboa Park and leverage the lower price to cross into Oakland.

The last two rows in table~\ref{tab:bart-arbitrage} are due to increased fares near the end of the routes shown in fig~\ref{fig:millbrae-richmond} and fig~\ref{fig:richmond-fremont}. This is similar to the red convex curve in fig~\ref{fig:dist-vs-fare}. So a commuter going from Richmond to Fremont can swap tickets with someone from Bay Fair and avoid the higher price of the full route. For a commuter from Bay Fair, Hayward and Fremont have very similar fares.

BART has 44 stations and ${44 \choose 2}$ = 946 unique trips. From these 946 trips we get $946 \choose 2$ = 446,985 unique pairs of trips. Out of these pairs of trips 60,334 or 13.5\% have arbitrage opportunity of at least 5 cents and 4,666 or 1\% have at least \$1 to be gained. The full list of these 4,666 pairs of trips with net arbitrage amount and percentage is available at this \href{https://github.com/asifhaque/transit-arbitrage/blob/master/data/sf/arbitrage.txt}{\texttt{github repo file}}\footnote{\href{https://github.com/asifhaque/transit-arbitrage/blob/master/data/sf/arbitrage.txt}{\texttt{github.com/asifhaque/transit-arbitrage/blob/master/data/sf/arbitrage.txt}}}.

Computing such pairs involves exhaustively finding overlapping paths between trips where the cost would be lower. Given that transit graphs are often spanning trees, even a brute force solution has complexity $O(n^5)$ where $n$ is the number of stations in the system. For BART $n$ = 44 and computation finishes in under a minute on a laptop. Source code, input and output data is available at \href{https://github.com/asifhaque/transit-arbitrage}{\texttt{github.com/asifhaque/transit-arbitrage}}

If BART authorities release anonymized data of every trip for users traveling through the system it would be very interesting to compute the total amount in dollars that San Francisco Bay Area commuters can save every day.

It would be interesting to explore Washington DC Metrorail in detail as well. Since it is a larger system with peak and off-peak rates, arbitrage strategies could be more complicated. Preliminary investigation exposes possibility of arbitrage.

\begin{figure}
\includegraphics[width=\linewidth]{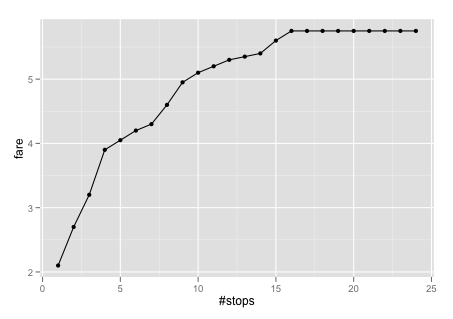}
\caption{Fare as a function of number of stops from Vienna/Fairfax station on the Orange line of Washington DC Metrorail.}
\label{fig:vienna-carrollton}
\end{figure}

Fig~\ref{fig:vienna-carrollton} shows fares from Vienna/Fairfax station towards New Carrollton on the Orange line of Washington DC Metrorail. We can see that the price structure is indeed very similar to the blue line in fig~\ref{fig:dist-vs-fare} and what we have seen for BART. The initial steep part passes through Falls Church suburbs leading to slight flattening around George Mason University (station 5). There is another steep part leading to DC city center around station 10. Heading out towards the suburbs prices are essentially flat for commuters coming all the way.

\section{Directions}
\label{sec:directions}

In this paper we have explored basics of transit fare arbitrage and analyzed BART fare structure both qualitatively and quantitatively via exhaustive computation. Our study highlights the need for careful design of pricing models for transit systems.

One interesting direction that might lead to an efficient fare system is if tickets are all electronic, ideally via a smartphone app, that third parties can manage to optimize inidividual trips. In recent years BART has moved from paper tickets towards Clipper Cards (NFC tags). Data is encrypted on these cards and can be read by any NFC reader but not written to except for BART machines. If in future we start using smartphone NFC capabilites (Google Wallet for example) and allow third parties to dynamically swap account profile on the smartphone app then arbitrage will be technically very easy. It may even allow arbitrage between routes that do not have overlapping paths and ultimately force the prices to be arbitrage free and thus efficient. We can imagine companies like Uber and Lyft managing such apps to provide commuters with a uniform payment interface encompassing ride shares and mass transit.

\end{document}